\documentclass[11pt]{article}
\usepackage{moriond,epsfig}

\newcommand{\Dzero}{D\O{} }
\newcommand{\etal}{{\em et al. }}
\newcommand{\Dyrad}{D{\sc yrad} }
\newcommand{\slE}{\mbox{/}\!\!\!\!E}
\newcommand{\GeV}{\mbox{ GeV}}

\begin{document}
\hfill Cavendish--HEP--98/09
\vspace*{4cm}
\title{What does the $W$ $p_T$ distribution tell us about the\\ $W +
\mbox{1 jet}/ W + \mbox{0 jet}$ ratio at the Tevatron ?}

\author{ D.J. SUMMERS }

\address{High Energy Physics Group, Cavendish Laboratory,\\
         Madingley Road, Cambridge CB3 0HE, England}

\maketitle
\begin{center}
\psfig{figure=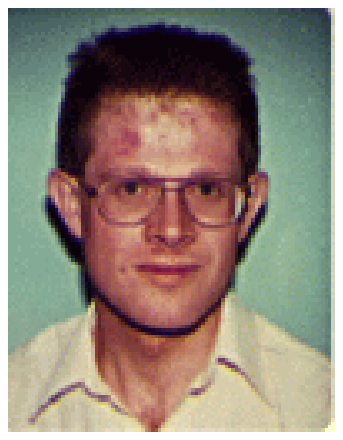,height=35mm}
\end{center}
\vspace*{5mm}
\abstracts{
We show that the $W$ $p_T$ distribution  measured by \Dzero at the
Tevatron agrees well
with the NLO QCD theoretical prediction for this quantity, whilst the
$W + \mbox{1 jet}/ W + \mbox{0 jet}$ ratio, ${\cal R}^{10}$,  measured
by \Dzero lies significantly above the NLO QCD theoretical 
prediction. We derive an approximate relationship between these two quantities,
and show that this rules out the majority of theoretical explanations
for the \Dzero excess in ${\cal R}^{10}$. We discuss possible
physics that could give rise to the ${\cal R}^{10}$ excess, which have
little effect on the $W$ $p_T$.}

% \section{Guidelines}
% \subsection{Producing the Hard Copy}\label{subsec:prod}

\begin{figure}[t]
\begin{center}
\psfig{figure=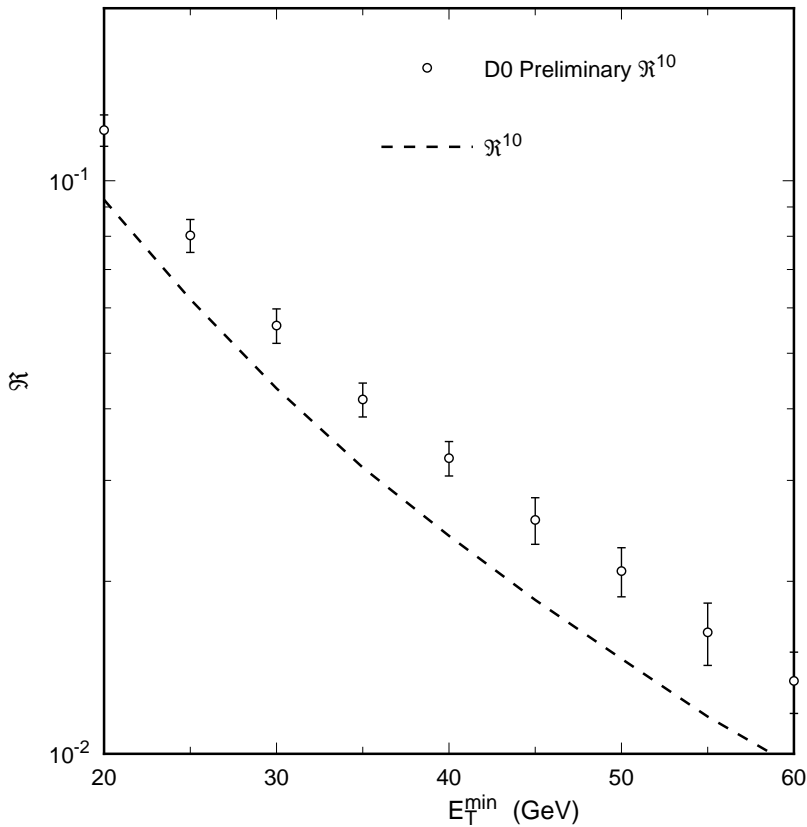,height=12cm}
\end{center}
\caption{\Dzero experimental measurement for ${\cal R}^{10}$,  
and the corresponding next-to-leading order QCD predictions. 
For the theoretical calculations we have chosen the scale $\mu=M_W$.
\label{fig:fig1}}
\end{figure}

For some time now \Dzero at the Tevatron has reported an excess in the
preliminary measurement for the ratio of 
$W + \mbox{1 jet}$ events to $W + \mbox{0 jet}$ events \cite{dzerorten},
\begin{equation}
{\cal R}^{10}(E_T^{\rm min}) 
={\sigma (W + \mbox{1 jet} ) \over  \sigma (W + \mbox{0 jet} )}\;,
\end{equation}
over the theoretical NLO prediction \cite{GGK},
where the jets are defined with transverse energy above some $E_T^{\rm
min}$, and both the numerator and denominator are exclusive with
respect to the number of jets. This is shown in Fig.\ref{fig:fig1}
where the preliminary \Dzero 
measurement lies about 30\% above the theoretical prediction for all
values of $E_T^{\rm min}$.

For this measurement $W$ bosons are tagged (both theoretically and
experimentally) in their leptonic decay to electrons by requiring
there to be an isolated lepton and significant amounts of missing
energy,
\begin{eqnarray}
E_T^e > 25\GeV, \qquad  |\eta^e| < 1.1,  \qquad \slE_T > 25\GeV,\nonumber\\
\Delta R (e,\mbox{jet}) > 0.4 \qquad {\rm for} \qquad E_T^{\rm jet} > 10 \GeV.
\end{eqnarray}
Then jets are formed in the rapidity range $|\eta_{\rm jet}|< 3.5 $
using the standard cone algorithm where all 
particles are clustered within $\Delta R$ cones, with $\Delta R =0.7$
\cite{HMPC,Jaethesis}.
For the theoretical predictions we simulate the experimental jet
algorithm by clustering
all pairs of partons that lie within $R_{\rm sep}\Delta R$ of each
other to form a proto jet, then test that all clustered partons lie
within $\Delta R$ of the proto jet \cite{Rsep}.
As a default
parameter, we set $R_{\rm sep} = 1.3$.  
The jet direction and transverse energy is constructed 
using the \Dzero recombination procedure \cite{D0recom}.

Now the definition for ${\cal R}^{10}$ can be rearranged to read,
\begin{equation}
{\cal R}^{10}(E_T^{\rm min}) =
{\displaystyle   \int_{E_T^{\rm min}}^\infty d E_T {1\over\sigma}\, 
                                {d\sigma^{\rm excl}\over dE_T}   \over
1 - \displaystyle \int_{E_T^{\rm min}}^\infty dE_T   
                                  {1\over\sigma}\, 
                                {d\sigma^{\rm incl}\over dE_T} } \;,
\end{equation}
% where $\sigma^{\rm excl,incl}$ are respectively the $W$ + 1 jet exclusive
% and inclusive rate. 
In this form the total $W$ cross-section, $\sigma$,
has no dependence on the value of $E_T^{\rm min}$ at which jets are
defined, and this means that the theoretical calculation  does not
contain any large logarithms of $E_T^{\rm min}$ and so should be accurately
calculable. We calculate $\sigma$
at fixed NLO in QCD\cite{GS}, that is  
${\cal O} (\alpha_S)$, using the program \Dyrad \cite{GGK}. We choose
to set the renormalization and factorization scale
set equal at $\mu=\mu_R=\mu_F=m_W$. On the other hand 
$d\sigma^{\rm incl,excl}/dE_T$, the  $W + \mbox{1 jet}$ rate,
inclusive or exclusive in the number of jets, which
explicitly depends on $E_T^{\rm min}$, we calculate at one higher order
in perturbation theory, that is NLO (as for finite $p_T$ we are forced
to have an additional parton), or ${\cal O} (\alpha_S^2)$. We keep the
renormalization and factorization scale equal, and by default use
$\mu=\mu_R=\mu_F=m_W$, although the theoretical calculation changes by
only a few percent if the scale is changed between 
$\mu=2m_W, m_W/2, E_T^{\rm jet}, \mbox{and } E_T^W$.
Throughout  we use the CTEQ4M parton densities 
\cite{CTEQ4} with $\alpha_s(M_Z)=0.116$.

This apparent difference between the experimental measurement and
theoretical prediction has led to several suggested explanations.
For example, Bal\'azs and
Yuan \cite{BY} have considered the effect of soft gluon resummation on
the related quantity ${\cal R}^{W}$,
\begin{equation}
{\cal R}^{W}(p_T^{W,\rm min}) = 
{\displaystyle   \int_{p_T^{W,\rm min}}^\infty d p_T^W {1\over\sigma}\, 
                                {d\sigma\over dp_T^W}   \over
1 - \displaystyle \int_{p_T^{W,\rm min}}^\infty dp_T^W   
                                  {1\over\sigma}\, 
                                {d\sigma\over dp_T^W} } \;.
\label{eq:rw}
\end{equation}
New physics effects are also possible and Choudhury \etal \cite{chou}
have considered the effect that a massive vector boson with the
quantum numbers of both a $W$ boson and a gluon would have on the
observed value of ${\cal R}^{10}$.  A more mundane explanation is that
an increase in the gluon parton distribution at medium Bjorken $x$
values would boost the $W$ +~1 jet rate, which receives contributions
from $qg$ scattering, while having little effect on the zero jet rate
\cite{dzerorten}.

In order to consider possible origins for difference between \Dzero's
measurement of ${\cal R}^{10}$ and the theoretical prediction of \Dyrad
it is worth considering variables that contain similar physical
information. At lowest order the observed jet in $W+\mbox{1 jet}$
events is produced by a single parton that recoils against the $W$
boson. This means that,
\begin{equation}
E_T^{\rm jet} = p_T^W\;.
\end{equation}
Beyond lowest order this equality is not exactly satisfied, however we
expect it to hold approximately. Hence we expect,
\begin{equation}
{\cal R}^{W}(p_T^{W,\rm min}) \simeq {\cal R}^{10}(E_T^{\rm min} 
= p_T^{W,\rm min}) \;.
\end{equation}

\begin{figure}[t]
\begin{center}
\psfig{figure=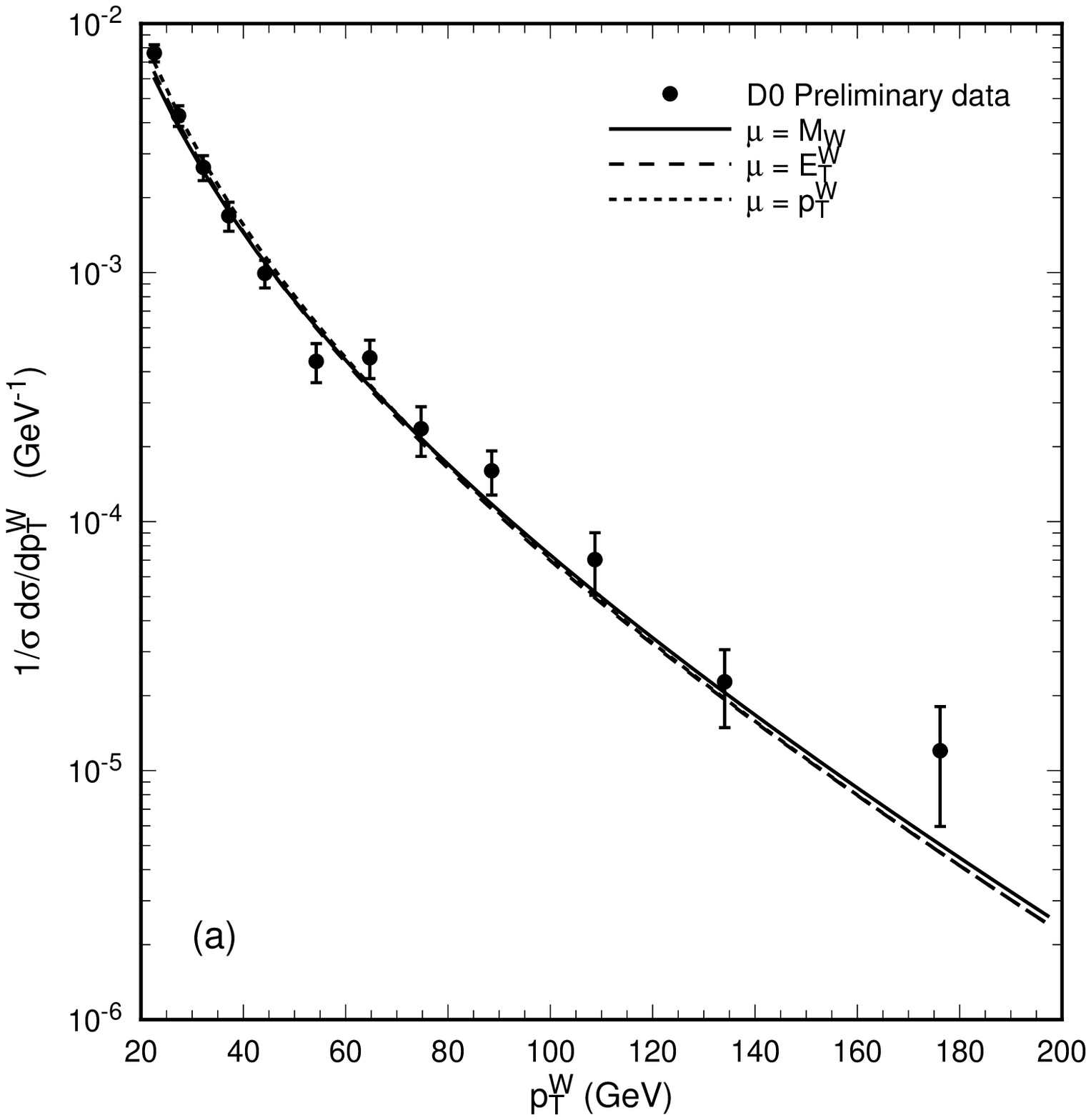,height=3in}
\psfig{figure=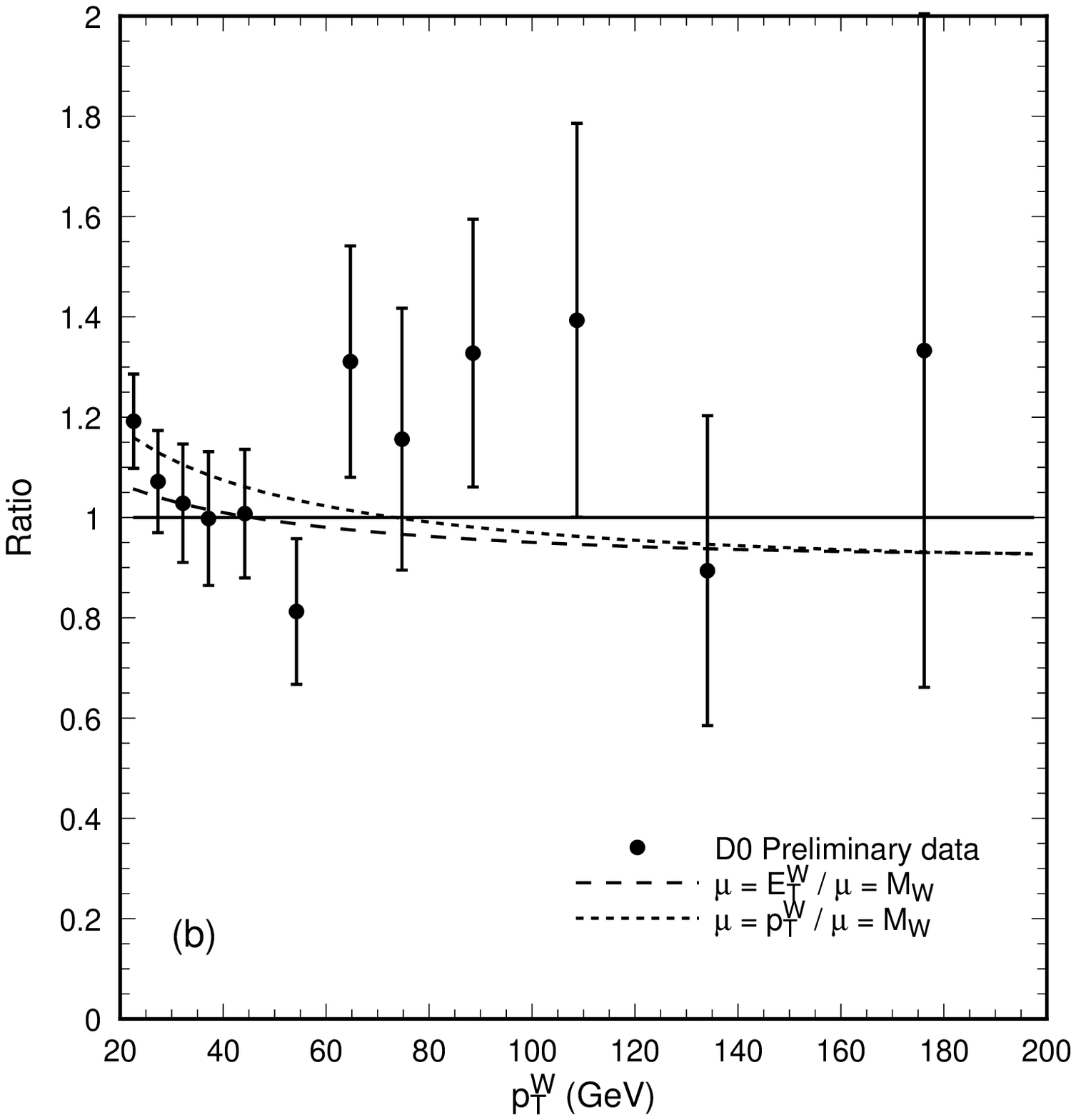,height=3in}
\end{center}
\caption{a) The \Dzero measured $p_T^W$ distribution, and the 
  ${\cal O}(\alpha_s^2)$ 
  predictions for the three choices of scale $\mu=M_W ,~E_T^W ,~p_T^W$;
  with the cuts described in the text.
  b) The ratio of the theoretical predictions, and the preliminary
  \Dzero measurement, to the theoretical prediction with $\mu=M_W$.
Note that in comparing with the \Dzero data, we have integrated over 
the appropriate range of $p_T^W$.
\label{fig:fig2}}
\end{figure}

Unfortunately ${\cal R}^W$ is not directly measured by \Dzero, however
they do measure the normalized $W$ $p_T$ distribution, 
$ {1/\sigma}\,{d\sigma/ dp_T^W}$. This is shown in
Fig.\ref{fig:fig2}, along with the NLO theoretical prediction for the
quantity. Clearly the same theoretical description that provided a bad
description of the ${\cal R}^{10}$ data, gives a good description of
the $W$ $p_T$ distribution.

\begin{figure}[t]
\begin{center}
\psfig{figure=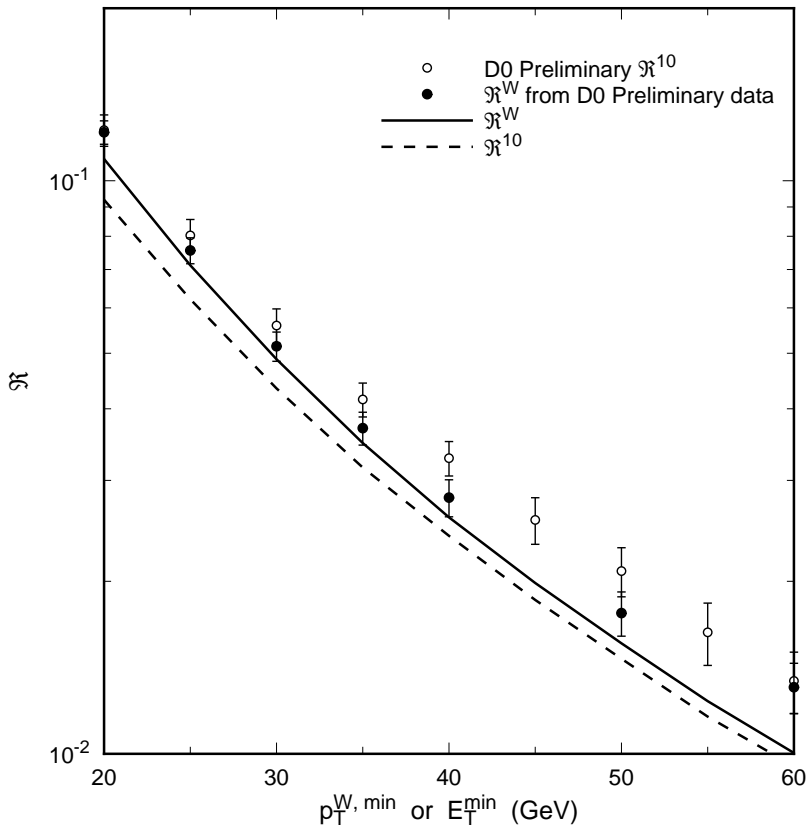,height=12cm}
\end{center}
\caption{Experimental measurements for ${\cal R}^{W}$ and ${\cal R}^{10}$, 
and the corresponding next-to-leading order QCD predictions. 
For the theoretical calculations we have chosen the scale $\mu=M_W$.
\label{fig:fig3}}
\end{figure}

The difference between the measurements of $W$ $p_T$ and ${\cal
R}^{10}$ can be made more directly, by transforming the $W$ $p_T$ into
${\cal R}^W$ using Eqn.\ref{eq:rw}. If we assume that the experimental
errors are independent this gives the result shown in Fig.\ref{fig:fig3}.
We also show the ${\cal R}^{10}$ measurement from Fig.\ref{fig:fig1}.

We can see that the theoretical prediction and experimental
measurement for ${\cal R}^W$ agree within errors, recall that due to
the definition of ${\cal R}^W$ each data point is not independent of
the others. That the theoretical prediction for ${\cal R}^W$ is
greater than ${\cal R}^{10}$ is easily understood, this is because 
${\cal R}^{10}$ is defined in terms of the exclusive 1 jet rate,
whereas ${\cal R}^W$ receives contributions from all $W$ events. This
means that if 2 or more jets are observed ${\cal R}^W$ receives a
contribution that ${\cal R}^{10}$  does not. Experimentally the
measurement for ${\cal R}^{10}$ lies above ${\cal R}^W$, which seems
hard to understand. This rules out most explanations for the measured
${\cal R}^{10}$ excess, as whatever explains the ${\cal R}^{10}$
excess must leave ${\cal R}^W$ unchanged.

\begin{figure}[t]
\begin{center}
\psfig{figure=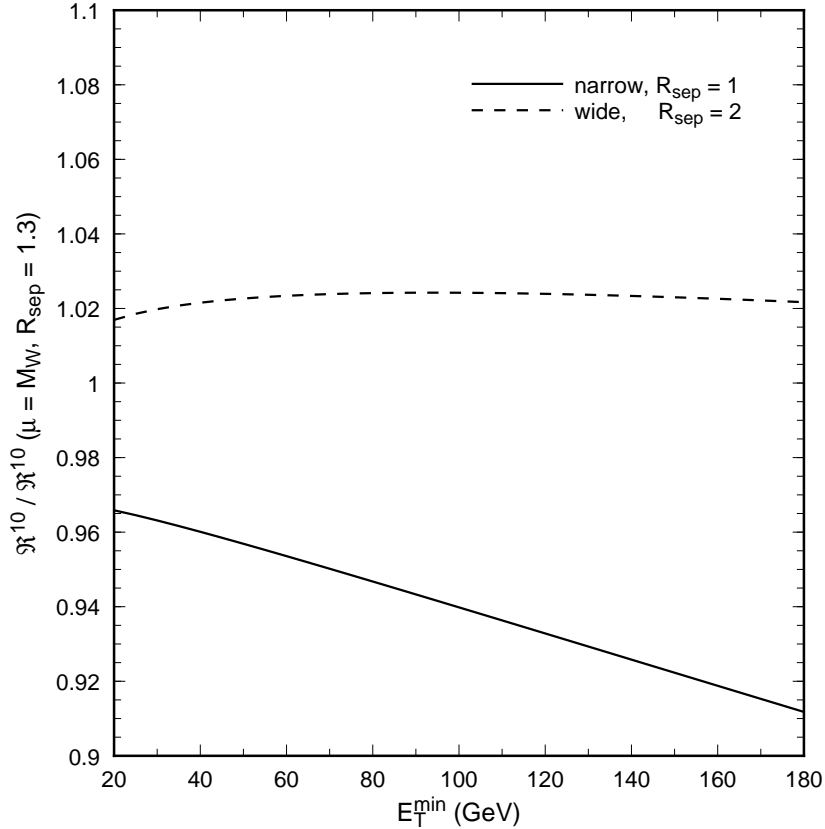,height=12cm}
\end{center}
\caption{The dependence of the \Dyrad prediction for ${\cal R}^{10}$ 
  on the jet
  clustering algorithm. We show the predictions
  normalized to that for $R_{\rm sep}=1.3$ clustering. 
  In all cases we have chosen $\mu = M_W$.
\label{fig:fig4}}
\end{figure}

So how can we understand the ${\cal R}^{10}$ excess in light of the
agreement between theory and experiment for ${\cal R}^W$? The
essential difference between ${\cal R}^{10}$ and ${\cal R}^W$ is that
for the former measurement jets need to be formed, whereas for the
latter they do not. At leading order the theoretical calculation in
insensitive to how jets are formed, however at NLO we gain sensitivity
to how jets are formed as two partons can by clustered into a single
jet; however this is still very far from the experimental situation
where typically many hadrons are clustered into each jet. Some feeling
for the difference between experiment and theory can be obtained by
varying the theoretical parameter $R_{\rm sep}$ between its natural
limits $ 1 < R_{\rm sep} <2$. $R_{\rm sep}=1$ corresponds
experimentally to there being no hadrons (seed towers) between the two
parton directions, while $R_{\rm sep}=1$ corresponds to their being a
seed hadron precisely between the two partons. We show the dependence
of the theoretical prediction on the theoretical parameter 
$R_{\rm sep}$ in Fig.\ref{fig:fig4}. Clearly this only changes the
calculation for ${\cal R}^{10}$ by a few percent, and so we do not
expect the difference between experimental and theoretical jets to be
the source of the difference between the QCD theory and experimental
measurement for ${\cal R}^{10}$.

That 
${\cal R}^{10}_{\rm Exp} > {\cal R}^{10}_{\rm Th}$ means that
experimentally there must be more energy associated with the jet than
theoretically. However the $W$ $p_T$ is not increased by this
additional energy clustered with the jet, this can only be if there is
even more additional energy flowing in the direction of the $W$ $p_T$
which balances the $p_T$ clustered with the jet. Hence to
simultaneously explain the ${\cal R}^{10}$ excess, while keeping
${\cal R}^W$ unchanged, one needs significant additional transverse
energy flowing in all directions. 

What possible explanations can there be for this additional energy?
Several ideas come to mind,
\begin{itemize}
\item soft gluons at higher orders in perturbation theory.
\item the underlying event.
\item multiple interactions.
\item overlapping events.
\end{itemize}
Each of these give additional energy in the event that will increase
any measured jets $E_T$, while the first two will only have a minor
effect on the $W$ $p_T$ and the last two have no effect on the $W$
$p_T$.

That theory and experiment are different for ${\cal R}^{10}$ is not a
sign that QCD is breaking down, as theory and experiment agree so well
for ${\cal R}^W$. Instead we should look at the differences between 
${\cal R}^{10}$ and ${\cal R}^W$ (or $ {1/\sigma}\,{d\sigma/ dE_T}$
and $ {1/\sigma}\,{d\sigma/ dp_T^W}$) as a probe as to how jets are
experimentally and theoretically formed. In this way $W$ events give a
somewhat independent test of how we study jet physics. $W$ events
through the production of the $W$ boson are known to have a hard
scattering, and this gives a somewhat different environment from the
usual environment 
in which jets are formed, and as such have somewhat orthogonal
sensitive to physics that can affect jets, such as the underlying
event and multiple interactions.

Finally we should note that the CDF collaboration at the Tevatron has
recently made their own measurement for the variable 
${\cal R}^{10}$,\cite{CDF} which agrees well with the NLO QCD theoretical
prediction \cite{GGK}. Although one may take this as a hint of an experimental
problem in the \Dzero measurement, this is far from clear as the
measurement made by CDF is of a slightly different quantity than \Dzero
measurement. CDF calculate the ratio of the inclusive 1 or more jet rate
to the inclusive zero or more jet rate, whereas \Dzero measure the
exclusive jet rates. As the fraction of $W$ events that contain two or
more jets is relatively small we do not expect the theory prediction to
work well in one case, but not in the other. Perhaps more importantly
CDF define their jets with a cone size of $\Delta R = 0.4$, and this
can have less innocent effects. For example if the \Dzero excess is
caused by a misunderstanding of the underlying event in $W$ events,
then as the underlying event is approximately flat in rapidity and
azimuthal angle, we would expect the larger \Dzero jet cones to show
approximately 3 times the effect of the smaller CDF cones. Such an
effect may cause the \Dzero measurement to be inconsistent with NLO
QCD theory, while the CDF measurement remains consistent.

%\newpage

\end{document}